\documentclass[useAMS,usenatbib]{mn2e}
\usepackage{graphicx}
\DeclareGraphicsExtensions{.png}
\pdfoutput=1

\title{The true stellar parameters of the Kepler target list}
\author[R. Farmer et al]{R.~Farmer\thanks{E-mail:
r.farmer@open.ac.uk}, U.~Kolb, A.J.~Norton \\
Department of Physical Sciences, The Open University, Walton Hall, Milton Keynes
MK7 6AA}

\begin{document}

\maketitle

\begin{abstract}
Using population synthesis tools we create a synthetic Kepler Input Catalogue (KIC) and subject it to the
Kepler Stellar Classification Program (SCP) method for determining stellar parameters such as
the effective temperature $T_{eff}$ and surface gravity $g$.
We achieve a satisfactory match between the synthetic KIC and the real KIC in the
$\log g$ vs $\log T_{eff}$ diagram, while there is a significant difference between
the actual physical stellar parameters and those derived by the SCP of the stars in the synthetic sample.
We find a median difference $\Delta T_{eff}=+500$~K and $\sim\Delta \log g =-0.2$~dex for main-sequence stars, and $\sim\Delta T_{eff}=+50$~K and $\Delta \log g =-0.5$~dex for giants, although there is a large variation across parameter space.
For a MS star the median difference in $g$ would equate to a $\sim 3\%$ increase in stellar radius and a consequent $\sim3\%$ overestimate of the radius for any transiting exoplanet. We find no significant difference between $\Delta T_{eff}$ and $\Delta \log g$ for single stars and the primary star in a binary system.
We also re-created the Kepler target selection method and found that the binary fraction is unchanged by the target selection. Binaries are selected in similar proportions to single star systems; the fraction of MS dwarfs in the sample increases from about 75\% to 80\%,
and the giant star fraction decreases from 25\% to 20\%. 
\end{abstract}

\section{Introduction}
The NASA Kepler mission \citep{borucki10} is designed to detect transiting
exo-earths in habitable zones around solar-like stars. To achieve this goal Kepler
is monitoring about 150,000 stars for 3 or more years.
The target stars were selected from a larger list, the Kepler Input Catalogue (KIC),
according to a set of criteria that rank stars in order of the likelihood to display
detectable transits of exo-earths in the habitable zone \citep{batalha10}. The KIC covers the 116 square
degrees of the Kepler field \citep{koch10} and contains about 450,000 stars with
magnitude brighter than $K_p = 16$ (where $K_p$ is the magnitude in the Kepler band). 
This catalogue was established to derive physical
parameters for objects in Kepler's field of view and to allow the selection of a
set of optimal targets that would maximise Kepler's chance of detecting an Earth-sized transit around a Sun-like star \citep{brown11}. The KIC itself was compiled 
from a ground-based survey using broad-band Sloan Digital Sky Survey (SDSS) filters with a flux precision of 2\%.

Kepler's Stellar Classification Program (SCP) \citep{brown11} derived basic
physical parameters of all KIC stars, chiefly the effective temperature $T_{eff}$,
surface gravity $g$, and metallicity $Z$, and, by comparison with suitable stellar
models, the stellar mass, radius and age, using only the observed broad-band
magnitudes and colours of these stars as an input. The target selection in turn
is based on these SCP-derived stellar parameters.

These SCP-derived parameters may suffer from random and systematic uncertainties
introduced because the measured magnitudes of a star may differ from its true,
intrinsic magnitudes, and because colours alone will not always unambiguously
deliver appropriate estimates of the physical parameters.
This will in turn translate into a bias of the statistical properties
of samples drawn from Kepler data, including the exoplanet candidate sample itself,
or the sample of binary stars with Kepler light curves.
 We note that stellar parameters are also needed to derive the properties of any
transiting planet that is detected \citep*{torress08}, but for confirmed planets the SCP
parameters are unlikely to be the sole or main source for the stellar parameters.

It is therefore important to critically examine the performance of the SCP approach,
and the consequences of any inherent systematic bias for the actual Kepler target list,
and for subsamples created from Kepler data.
To this end we aim to create a synthetic version of the KIC, obtained by population synthesis calculations that include self-consistently evolved binary systems.
We validate the population model against the actual KIC in colour-magnitude
space, and employ the SCP technique to derive "apparent" stellar parameters for all stars in the synthetic sample, i.e. exclusively from their magnitudes in different colour bands.
We then investigate the difference between the actual, physical parameters of our synthetic stars, and their SCP-derived parameters.

Due to bandwidth limitations Kepler does not
observe every object in the KIC, instead a target list is drawn up that aims to maximise
the science return on the targets observed. This list is determined on the basis of the SCP-derived
parameters and the expected flux levels, aiming to increase the
fraction of Sun-like stars and decrease the fraction of giants in the sample. In this study we wish to reproduce the target selection procedure and apply it on the synthetic sample of the KIC, to quantify the resulting bias against giants on the basis of the actual, physical parameters of the synthetic KIC stars.

The paper is organised as follows. In Section \ref{sec:method} we describe our population synthesis
model, including updates we made to the existing BiSEPS code. Section \ref{sec:kic} deals with the derivation of the Kepler
target list and a comparison of our work to the real KIC. In Section~\ref{sec:results} we compare the real physical parameters of our synthetic sample to those derived from the SCP. We then investigate the bias introduced by the target selection method. In Section~\ref{sec:discuss} we discuss the significance of various assumptions made in our analysis, while in Section~\ref{sec:con} we summarise our main findings.

\section{Population synthesis model}\label{sec:method}

To calculate a model for the stellar and binary star population in the
Kepler field-of-view we added new input physics and functionality to the
BInary and Stellar Evolution Population Synthesis (BiSEPS) code  which was
originally described in
\citet{willems02,willems04} and later employed by \citet{willems06} in a simplified way to study
the false positive rate in the exoplanet transit search project SuperWASP
\citep{pollacco06} from shallow-eclipsing binaries.  BiSEPS in turn is based
on the analytical descriptions of stellar and binary evolution by
\citet*{hurley00} and \citet*{hurley02}.

\subsection{Binary evolution}\label{sec:biseps}

At the core of the population synthesis scheme is a large library of single star and binary system evolutionary tracks from the ZAMS up to a maximum age of 13 Gyrs, providing physical parameters for typically 100 time steps suitably distributed along the tracks.
The stellar evolution scheme takes into account mass loss via winds, Roche lobe overflow, and angular momentum losses due to gravitational wave radiation and magnetic braking (see \citet{willems04} for references).
A newly forming binary system is taken to be fully characterised by the initial masses of its components, the orbital separation, and the stars' chemical composition, set here with hydrogen abundance $X=0.70$ and metallicity $Z$ (we consider either $Z=0.020$ and $Z=0.0033$).
All systems start with and are forced to have, during their evolution, circular orbits.

The initial parameter space is divided into 50 logarithmically spaced equidistant bins of initial masses
$M_1$ and $M_2$ between $0.1$ and $20M_{\sun}$
and into 250 logarithmically spaced equidistant bins of initial semi-major axes $a$ between
3$R_\odot$ and
$10^6 R_\odot$. By symmetry, only objects where $M_1 \gid M_2$ are evolved. Single star tracks are obtained from the primary star tracks in very wide, non-interacting binaries (with $a=10^7 R_\odot$).

\subsection{Galactic Model}
Underpinning the spatial distribution of the synthetic stars is a simple kinematic model of the Galaxy,  described in detail in
\citet*{willems06} (and references therein).
The Galaxy is assumed to comprise a young thin
disc and an older thick disc. Each disc's stellar distribution is modelled as a
double exponential of the form
\begin{equation}\label{eq:stelden}
\Omega(R,z)=n_o \exp \left(\frac{-R}{h_R}\right)\exp
\left(\frac{-|z|}{h_z}\right)
\end{equation}
with $h_R=2.5$\ kpc and $h_z=300$\ pc for the thin disc and
$h_R=3.8$~kpc, $h_z=1$\ kpc for the thick disc . The integral is normalised to unity,
thus $n_o=1/4\pi h_R^2 h_z$.
We assume that star formation proceeded for the first 3 Gyr after the formation of the Galaxy in the thick disc and then continued until the current epoch (13 Gyr) in the thin disc.
During the respective star forming periods the star formation rate is taken to be constant in each disk, such that one star or binary with component mass $M>0.8M_{\sun}$ is produced per year \citep{weidemann90}.
To capture the essence of the metallicity evolution with Galactic age we go beyond \citet{willems06}
and assume that thick disc stars have a metallicity
$Z=0.0033$ \citep*{gilmore93}
while stars forming in the thin disc have a solar metallicity value of $Z=0.020$ \citep{haywood01}.

To obtain the total number of systems in a given survey field the stellar density as
defined in (\ref{eq:stelden}) is numerically integrated over Galactic longitude,
latitude and distance ($l,b,d$) by translating it from
galactocentric ($R,z$) to heliocentric coordinates ($l,b,d$) via
\[
R=\left(d^2\cos^2b -2dR_{\sun}\cos b \cos l
+R^2_{\sun}\right)^{\frac{1}{2}}
\]

\begin{equation}
z=d\sin b + z_{\sun},
\end{equation}
where $R_{\sun}=8.5$kpc is the radial distance of the Sun from the Galactic
centre \citep{reid93} and
$z_{\sun}=30$pc is the height of the Sun above the Galactic plane
\citep{chen01}.

For each system the integral over distance is carried out between the minimum and maximum
distance, $d_{min}$ and $d_{max}$, this system can be seen at. If the survey is magnitude-limited these are determined by
$d=10^{\left(m-M+5-A_\lambda\right)/5}$ where $m$ is the lower or upper magnitude limit of the survey,
$M$ the absolute magnitude and $A_{\lambda}$ is the extinction along a line of sight at $(l,b)$ in the filter band of the survey.

The target magnitude range of the Kepler mission is $8 \lid K_p \lid 16$ \citep{batalha10}, and it is this interval we used for computing the synthetic KIC.
We extend our  simulations to include stars down to $K_p =19$ so
that we can estimate the background flux levels, and we model bright stars up to, arbitrarily, $K_p =0$, to include the few bright objects that will saturate the detector but not be observed as target objects.

As the extinction itself depends on the distance, we calculate the distance limits for the integral iteratively.
We upgraded the BiSEPS extinction routine described in \citet{willems06} to that of
\citet{drimmel03} which calculates the Galactic extinction from a 3D dust model of the Galaxy that has been scaled
using data from the COBE/DIRBE NIR instrument to provide extinction values along lines of sight in the \textit{V}-band (see also Section~\ref{sec:bol_corr} below).

\subsection{The Kepler field}

The integration boundaries for Galactic longitude and latitude are determined by the location of the
field-of-view of Kepler's 42 CCDs; each of these is split into two distinct regions along the channel boundary.

For practical reasons we define integration regions bounded by lines of constant $l$ and $b$ that enclose the individual Kepler
CCD areas, and split these regions up
into smaller boxes. The result of the volume integration for each of these boxes is then weighted proportional to the fraction of its area that overlaps with its respective CCD channel, using the ConvexIntersect routine from \citet{orourke93}.

The numerical volume integration for each box makes use of a Romberg integral following \citet{press92}.
This divides the volume up into at least $2^5$ intervals in each of the directions $l,b,d$, and iteratively
increases the number of intervals by factors of 2 up to a maximum of $2^{10}$
intervals, until the integral changes from one iteration to the next by less than 0.1\%. If this condition is not met
once $2^{10}$ intervals are reached the integral obtained for $2^{10}$ intervals is used.
We found that decreasing the cut-off to below 0.1\% did not significantly alter the results, while markedly increased the computational runtime. Very few integration areas ever needed more than $2^5$
intervals.

\subsection{Population characteristics}

The system-specific observable volumes
are then multiplied by weighting factors determined from the distribution functions of newly-formed
stars and binaries to calculate the total number of each type of binary and single star that are visible in the Kepler field.
We adopt an initial mass function (IMF) with a slope -1.23 for $0.1 M_{\sun} \lid M_1 < 0.5M_{\sun}$, -2.2 for
$0.5 M_{\sun} \lid M_1 < 1.0 M_{\sun}$ and -2.7 for $1.0 M_{\sun}\lid M_1$ \citep{kroupa01}, for both
single stars and the primary star of binary systems. The secondary mass is selected from a flat initial mass ratio distribution (IMRD).
The distribution of initial orbital separations is assumed to be $\chi(\log a)=0.078636$ for $3\lid a /R_{\sun}\lid
10^6$. The lower limit is a simplistic cut \citep{hurley02},
while binaries beyond the upper limit are likely to be disrupted by passing
intergalactic stars \citep{heggie75}.
Finally we assume that 50\% of all systems form as binaries.

\subsection{Bolometric corrections}\label{sec:bol_corr}

To expand the possible filter sets BiSEPS can deal with from
just the V band to the Johnson-Cousins-Glass \textit{UBVRIJHK}, Str\"omgren
\textit{uvby}$\beta$, Sloan
\textit{ugriz} and Kepler $K_p$ bands as well as the custom \textit{D51} band used by the SCP for the KIC
we have updated the bolometric corrections (BCs) from those given by \citet{flower96} to those
of \citet{girardi02}. The BCs are provided as a function of  $T_{eff}$ and $\log g$ in the form of tables for different metallicities. They are based on the synthetic ATLAS9 spectra for stars between
$3900 \rmn{K}< T_{eff}< 50000 \rmn{K}$ and $0< \log g < 5$ \citep*{castelli97}, and the
BDdusty1999 atmosphere models \citep{allard00} for stars with
$700 \rmn{K} < T_{eff} < 3900 \rmn{K}$; stars hotter than $T_{eff} >
50000 \rmn{K}$ are treated as black-bodies. M giants are treated separately
by using the empirical spectra of \citet{fluks94}. These stellar spectra are integrated over the filter response curve to derive the bolometric corrections for a star in any filter system \citep[see][for further details]{girardi02}.

We perform a bi-linear interpolation over
$T_{eff}$ and $\log g$ for tabulated metallicities either side of the target metallicity and then a linear interpolation between the two metallicities.
If the parameters of a star place it outside of the range provided by
the tables in \citet{girardi02} we use the closest point inside the tables, rather than risking extrapolating the data. With the BC defined for a specific
$T_{eff}$, $\log g$ and metallicity we then calculate the absolute magnitude $M_x$ of a star
for a specific filter $x$ as
\begin{equation}
M_x=-2.5\log \left(\frac{L}{L_{\sun}}\right)-M_{Bol,\sun}-BC_x
\end{equation}
where $L$ is the star's bolometric luminosity, as delivered by the evolutionary model, $M_{Bol,\sun}$ is the Sun's
bolometric magnitude and $BC_x$ is the bolometric correction for a star in
filter band $x$. We calculate $M_{Bol,\sun}$ in a self-consistent way from
\begin{equation}
M_{Bol,\sun}=M_{V,\sun}+BC_{V,\sun}
\end{equation}
We take the Sun to have $T_{eff}=5777 \rmn{K}$ and $\log g=4.44$,
giving $BC_{V,\sun}=-0.06$ \citep{girardi02}. Defining the visual apparent magnitude of the Sun to be $V_{\sun}=-26.76$ implies $M_{V,\sun}=4.81$, and hence
$M_{Bol,\sun}=4.75$ \citep[see][for a
review]{torres10}.

For the purposes of this work
we follow the SCP and use a combination of the $g$ and $r$ band magnitudes to derive the $K_p$ magnitude, using
equations 2a and 2b from \citet{brown11},
\[
 K_p = 0.1 \times g+0.9 \times r \quad  \rmn{for} \quad (g-r)\leq0.8
\]
\begin{equation}\label{eq:kepmag}
K_p= 0.2 \times g+0.8 \times r \quad \rmn{for} \quad (g-r)> 0.8
\end{equation}

\subsection{Extinction}
We obtain the extinction $A_\lambda$ in a given filter band from the extinction $A_V$ in the visual band
calculated from \citet*{drimmel03} via the relation
$A_\lambda / A_V = \Lambda$, where $\Lambda$ is a filter dependant coefficient \citep{girardi08}.
For simplicity we follow the SCP approach and
adopt a single value of $\Lambda$ for all stars in each filter band, neglecting the real dependence on $T_{eff}, \log g, Z$ \citep{girardi02}. We chose the coefficients of a 5000~K, $\log g=4.0, \log (Z/Z_{\sun})=0.0$ star \citep{girardi02}, which are given in Table \ref{tab:extinct}.

\begin{table}
\centering
\begin{tabular}{|l|c|}
 \hline
 Filter & Extinction coefficient$\left(\Lambda\right)$ \\
 \hline
g & 1.193 \\
r & 0.868 \\
i & 0.681 \\
z & 0.490 \\
D51 & 0.999 \\
 \hline
\end{tabular}
\caption{Extinction coefficients for a 5000~K, $\log g=4.0, \log (Z/Z_{\sun})=0.0$ star, from tables provided in \citet{girardi02}}
\label{tab:extinct}
\end{table}

\subsection{Creating a discrete sample}
\label{sec:subsample}
The result of the above volume integration and weighting with initial distributions is a multi-dimensional, continuous (albeit binned) distribution function $\Gamma$ that characterizes the content of the Kepler field-of-view at the current epoch.
The total number $N$ of stars and binary systems is defined by the integral over these distributions.
To obtain a synthetic KIC which can then be subjected to the Kepler target selection procedure we create a discrete synthetic sample of $N$ stars from this continuous distribution.

To this end we draw a random sample of $N$ objects from the distribution function $\Gamma$.
Each object in the sample is placed randomly at a location ($l$, $b$, $d$) inside
the field-of-view, based on the Galactic density, extinction and absolute magnitude
of the system. We obtained a number of different random samples and found no significant difference in the sample properties discussed below.

\section{Kepler target list selection}\label{sec:kic}

Out of the possible 450,000 stars in the Kepler field, only $\sim 150,000$ can be observed at any one time due to bandwidth limitations.
Therefore Kepler uses a tailored target list selected according to a number of criteria designed to maximise the likelihood for the detection of Earth-like transits in the star's habitable zone \citep{batalha10}.
To be able to generate a synthetic target list from our synthetic sample that would reproduce the actual Kepler target list
we created our own model of the Kepler detector system and target selection method.
Following the procedures as set out in \citet{brown11}, \citet{bryson10} and \citet{batalha10} this entailed the following principal steps: (a) derive estimates of the system parameters from broad-band colours using the SCP routine, (b) construct a model of the expected S/N measured by each pixel, and then combine (a) and (b) to calculate the likelihood of detecting Earth-like transits in the star's habitable zone.

In essence, to compile the target list the stars are ranked in terms of the minimum radius $R_{p,min}$ of a planet that can still be detected securely in the absence of intrinsic stellar noise within the 3.5 yr mission. 
The radius $R_{p,min}$ is obtained by requiring that the relative transit depth in flux $F$, $\Delta F/F = (R_p/R_*)^2$, where $R_*$ is the stellar radius, exceeds a suitable multiple of the light curve noise $\sigma_{tot}$. This becomes 
 \begin{equation}\label{eq:target}
R_{p,min}=R_*\sqrt{\frac{7.1\sigma_{tot}}{r}}
\end{equation}
\citep[Equation 7 of][]{batalha10}
where $r$ is a crowding metric and discussed below. 
Choosing to set the noise level to $7.1\sigma$ also implies that there would only 
be one statistical false positive signal due to random fluctuations in the light curve \citep{batalha10}. 

We now discuss the different factors in equation~\ref{eq:target}.

\subsection{Stellar classification}\label{sec:stellClass}

The determination of physical parameters of all KIC stars, including the stellar radius $R_*$, is the remit of the SCP. 
This uses a Bayesian posterior probability estimation method to derive a star's $T_{eff}, \log g, \log Z$, luminosity, mass and radius from its observed colours \citep{brown11}.

The two-step procedure is based on two sets of input models. Stellar atmosphere models of \citet{castelli04} were used with filter response functions to determine the expected colours for objects between $3500 \rmn{K}< T_{eff}< 50000 \rmn{K}$, $0< \log g < 5.5$ and $-3.5< \log (Z/Z_{\sun}) < 0.5$ (although not every gravity is available at every temperature), while stellar evolution tracks of \citet{girardi00}, assuming a constant star formation rate and solar metallicity, link these with the stellar mass and radius. 

Bayesian priors based on the $T_{eff}, \log g$ distributions of stars observed
by the \textit{Hipparcos} \citep{perryman97} satellite, the $\log Z$ distribution from \citet{nordstrom04} and a Galactic distribution model from \citet[][pg482]{cox00} are employed to focus the search in parameter space. The claimed advantages of a Bayesian approach is that a prior rules out implausible systems which e.g.\ a standard $\chi^2$ minimisation technique might obtain. However shortcomings were noted in \citet{brown11}; the metallicity distribution was deemed questionable, $T_{eff}$ is unreliable for the hottest and coolest objects and there are systematic errors in $\log g$ for objects with $g-r>0.65$.

For each object in our synthetic sample we supply the calculated \textit{g r i z} and \textit{D51} magnitudes as an input for the SCP code, to estimate the object's physical parameters in the same way as the SCP did for the stars in the real KIC\footnote{http://www.cfa.harvard.edu/kepler/kic/kicindex.html} \citep{brown11}.
The SCP code takes into account magnitude uncertainties, and for simplicity we selected a value of 0.02 mags in each band for all stars, which is the quoted photometric
precision for objects with $K_p<15$, as measured by the SCP \citep{brown11}.
As the KIC required excessive exposure times in the \textit{u} band we excluded it from the fitting process by selecting a large photometric uncertainty for it.  We also found that the J, H and K magnitudes had little effect on the results, and thus excluded these bands as well, to reduce the number of unnecessary fit parameters and save CPU time (see section:\ref{sec:colcol}).

Binary stars were treated as point sources, with a magnitude in each filter band given by the sum of the fluxes of the two stars in that filter band.

\subsubsection{S/N determination} 
Determining the expected S/N for an observation requires knowledge of Kepler's noise characteristics a model for which exists in \citet{bryson10a}, however the tools required are not publicly available and therefore we re-derive them here. 

To calculate the S/N expected for each synthetic system from its $K_p$ magnitude and the system's RA and DEC we require a model of Kepler's focal plane geometry (FPG), as described in the following. 

To place the synthetic star on the focal plane we obtained its pixel coordinates by extrapolating those of the closest match in the actual Kepler data set, based purely on the star's RA and DEC\footnote{http://keplergo.arc.nasa.gov/ContributedSoftwarePyKEP.shtml}.
Stars near the centre of the field have almost circular pixel response functions (PRF), while near the edge the PRFs are elongated towards the centre of the field \citep{bryson10}. Thus the PRFs are both a function of CCD and pixel location of the system.
\citet{bryson10} defines a set of 5 PRFs for each CCD, four in the corners and one in the centre. Each of these PRFs gives the flux distribution over a $n \times n$ pixel array, with usually $n=11$ but occasionally $n=15$, for a star centred in the middle of the grid.
To derive the PRF of a synthetic system we linearly interpolate between the 2 nearest corner PRFs and the central PRF. 

In this way we build up 
a full frame image (FFI) of all synthetic stars in the Kepler field down to a limiting magnitude of $K_p=19$,
which was chosen as the assumed zodiacal light emission equates to a 19\textsuperscript{th} magnitude star on each pixel \citep{jenkins04}.

With the FFI in place we can determine the noise per pixel, as described in \citet{caldwell10} and summarised here.
For each synthetic star we calculate the PRF and subtract this from the FFI, to obtain an image
of the system on its own as well as of the background around the system, including the zodiacal light. We convert the flux to electrons via
\begin{equation}
f_{kep}=10^{-0.4(K_p-12)} \times f_{12}
\end{equation}
where $f_{12}=1.74 \times 10^5 e^-$s$^{-1}$ is the photoelectric signal for a G2 V star with $K_p=12$ \citep{jenkins10}.
We then apply a smearing to each image, by summing the flux of each pixel in each column, multiplying by the read time of 0.52s, dividing by the number of rows and adding this to each pixel.

At this point we apply a saturation model by `rolling over' the electrons which are above the well depth \citep{caldwell10}.
This is done by performing a 50/50 split of the overflowing electrons, moving half of them up the pixel column and half down the pixel column, with
each subsequent overflow moving electrons in the same direction; until such a point that the number of electrons per pixel is at most the well depth
\citep{cleve09}. A charge transfer efficiency model is then applied with a value of 0.99993 for the parallel reads and 0.99995 for the serial reads
\citep{cleve09}. 

With both images now expressed in electrons and the various systematics applied we calculate the S/N ratio for each pixel using 
\begin{equation}
S/N=\frac{S}{\sqrt{S+Bg+\sigma_{read}^2+\sigma_{quant}^2}}.
\end{equation}
In this version of the CCD equation the signal $S$ and background $Bg$ are given in electrons, while the read noise $\sigma_{read} \sim 100 e^-$ per read is CCD dependant \citep{cleve09} 
and the quantisation error $\sigma_{quant}$ is given by 
\begin{equation}
\sigma_{quant}=\sqrt{\frac{1}{12}}\left(\frac{\rmn{W}}{2^{N_{bits}-1}}\right).
\end{equation}
\citep{bryson10a}. Here the CCD well depth W is of order $\sim 10^6 e^-$ per pixel (though it is also CCD dependant, see \citet{cleve09}), and $N_{bits} = 14$ denotes the number of bits the data is quantised to. This gives $\sigma_{quant} \sim 30 e^-$ per pixel.
The pixels are ranked in order of decreasing S/N and summed in quadrature, until the sum of the S/N is maximised, thus defining the optimum aperture for the star. This is repeated for each star with $K_p \leq 16$.

The total photometric error $\sigma_{tot}$ is obtained from the S/N value, scaled by the total number of individual integrations while the system was in transit over the envisaged 3.5 years of the mission. 
This number is the product of the 270 integrations co-added together in one long-cadence (30 min) observation, the number $N_{sample}$ of long cadence observations that fit in a single transit, and the number $N_{tr}$ of transits in 3.5 years. We thus have 
\begin{equation}\label{eq:sigmatot}
\sigma_{tot} = \frac{N}{S \times \sqrt{270 N_{sample} N_{tr}}}.
\end{equation}
For randomly distributed inclinations of circular orbits the average transit duration is $t_0 \pi/4$, where $t_0$ is the duration of a transit that is central across the star. Thus we have $N_{sample} = t_0 (\pi/4) / 30$\ min.
The central transit duration is calculated as
\begin{equation}\label{eq:centTransit}
t_0 = 2R_*\sqrt{\frac{a}{GM_*}},
\end{equation}
with the stellar mass $M_*$ derived from the SCP, and with the semi-major axis $a$ taken at 
three different locations, 5$R_*$, 0.5$H_*$ and $H_*$. The quantity $H_*$ is the characteristic distance of the habitable zone (HZ) for the star in consideration and is given by $0.95\sqrt{L_*/L_\odot}$\citep{batalha10}.

The final term required for evaluating equation \ref{eq:target} is the crowding metric, $r$, which is given by \citep{batalha10}
\begin{equation}
r=\frac{F_{*}}{F_{*}+F_{bg}}
\end{equation}
where $F_{*}$
is the flux from the star in the optimal aperture before addition of the systematics, and $F_{bg}$ is the flux from the background in the optimal aperture before addition of the systematics but after the zodiacal light has been added.

\subsection{Testing the target selection code}

With $R_{p,min}$ calculated for each synthetic star in the field-of-view we can draw up a ranked list of stars in order of increasing $R_{p,min}$. The
subset of systems with a detectable terrestrial sized transit in the habitable zone, i.e. $R_{p,min} \leq2 R_E$ (where $R_E$ denotes the radius of the
Earth) includes
a large number of objects, $\sim60\%$, that are too faint ($K_p>15.0$) for radial velocity follow up. Thus an additional prioritisation scheme is
employed, the details of which are given in table 1 of \citet{batalha10}. In essence, the highest priority stars are those with $R_{p,min} \leq 2 R_E$
in the HZ, with a magnitude bright enough to perform high precision radial velocity on ($K_p\leq14$), followed by those with $14\leq K_p\leq16$. Then
there are those with detectable Earth-sized planets at $a=0.5H_*$ or $a=5R_*$ (these deliver a larger number of transits over the lifetime of the
mission), and finally those with $R_{p,min} <2R_e$ in the HZ around the faintest stars. \citet{batalha10} divides the sample into 13 classification
groups, with the 11 highest priority groups making up the target list.

To test the target selection code we apply it to the actual KIC and compare the target list we obtain with the Kepler Quarter 2 (Q2) data set which we use as a proxy for the actual Kepler target list.
We chose Q2 as the catalogue of objects defined in \citet{batalha10} is not publicly available, and because both  
Quarter 0 and Quarter 1 were affected by commissioning of the Kepler instrument.
In Q0 only $\sim 50,000$ stars were observed \citep{borucki11a}, while Q1 had over-sized apertures \citep{borucki11},
thus a reduction in the number of faint, $K_p=15-16$, stars. We chose not to use later quarters either 
because after each quarter some targets are removed due to follow-up work,
or added due to the guest observation program. \citet{tenenbaum12} shows that in the first 12 quarters,
60\% of objects were observed for all 12 quarters. A further 15\% were observed for 10 quarters; these predominately are
systems falling on the CCD module that failed during quarter 4, and thus were only observable for 75\% of the time.

\begin{figure}
\includegraphics[width=1.0\linewidth]{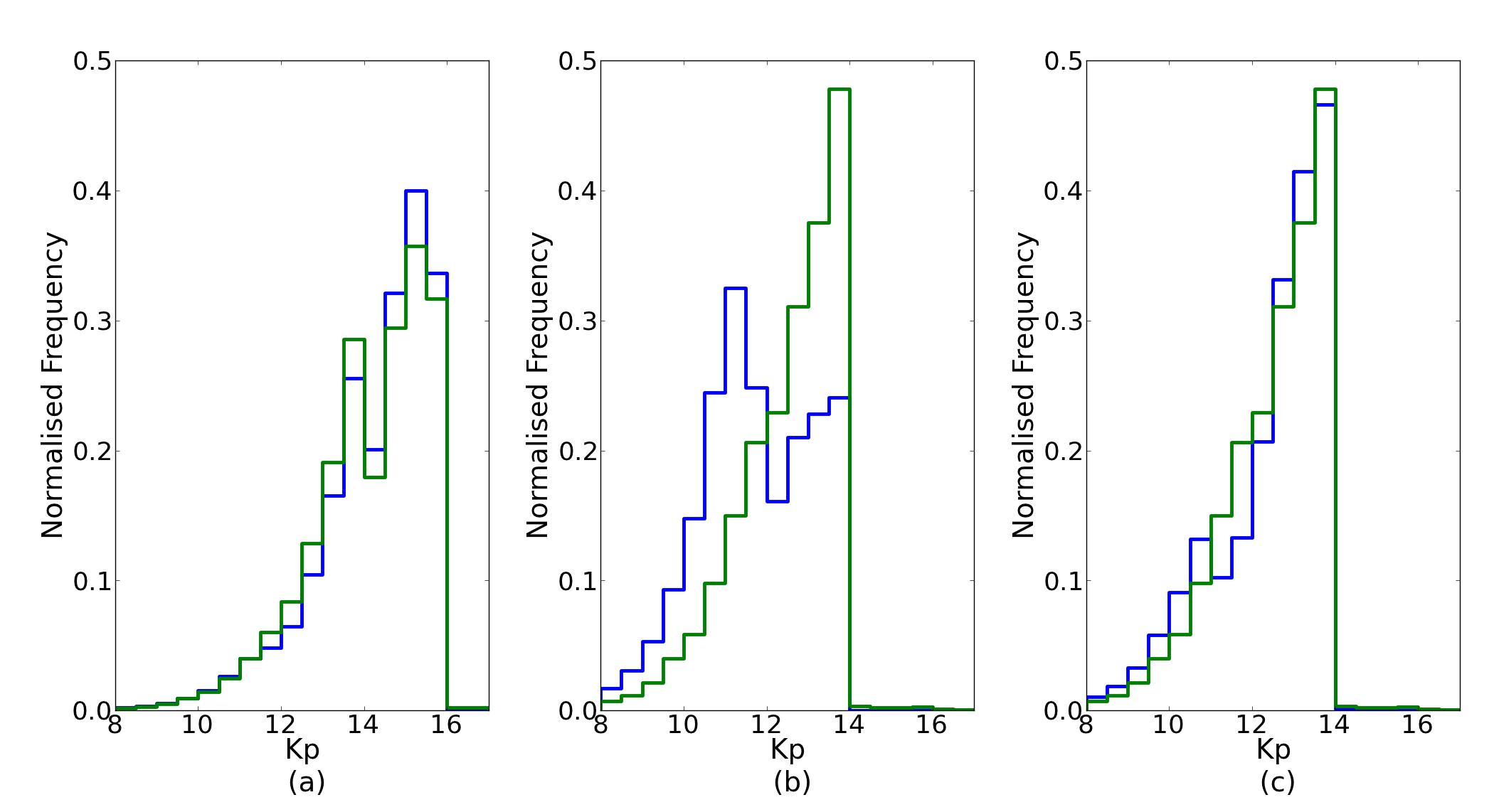}
\caption{\textit{Blue}: Normalised magnitude distribution of stars selected by our procedure. Left: all systems, no ad-hoc corrections. Middle: only giants, defined as objects with KIC $\log g <3.5$, no ad-hoc corrections. Right: only giants, but with ad-hoc corrections. \textit{Green}: corresponding sample from the actual Q2 target catalogue. Panel (a): all targets; panels (b) \& (c): only giants.}
\label{fig:adhoc}
\end{figure}

We show the magnitude distributions of our calculated target list and of the actual target list in Fig.~\ref{fig:adhoc}a. The discrepancy seen is primarily due to giant stars, here defined as stars with $\log g <3.5$, highlighted in Fig.~\ref{fig:adhoc}b.
We could not attribute these differences to inadequacies in our implementation of the target selection and SCP code and rather suspect that at least some differences exist because the actual Q2 list will have some objects added or removed from the original list of objects as defined in \citet{batalha10}.

To achieve a better agreement we applied a series of ad-hoc corrections to our target selection criteria:
\begin{enumerate}
\item For faint objects ($14<K_p<16$), if $R_{p,min} \leq 2.0 R_E$ for $a=H_*$, we redefine the selection criterion to $R_{p,min} \leq 2.4 R_E$. This increases the number of faint dwarfs.
\item All objects that saturate at least one pixel are included, if they have not
already been placed into one of the groups in \citet{batalha10}. This predominantly increases
the number of bright giants.
\item All objects with $3<R/R_{\odot}<10$ and magnitude $K_p<14$ are included, if they have not
already been placed into one of the groups in \citet{batalha10}. This is purely ad-hoc and is designed to increase the number of bright giants.
\end{enumerate}
With these corrections in place we consider the match between the reproduced and actual target list satisfactory (see Fig.~\ref{fig:adhoc}c) and sufficient for the study of system properties presented in the following sections.

\section{Results}\label{sec:results}
We now present the synthetic Kepler field population, covering both the synthetic KIC and the synthetic target catalogue which emerges from it. We will compare the actual physical properties of the synthetic stars with the properties these stars appear to have when analysed with the SCP method.

\subsection{Sample size}\label{sec:stelcal}
Using the default population synthesis parameters described above we obtain a total of $\sim353,000$ objects in the synthetic KIC, compared to the
$\sim416,000$ objects in the real KIC. Increasing the Galaxy-wide SFR from the default value of 1 star yr$^{-1}$ with $M >0.8M_{\odot} $ to 1.2 stars yr$^{-1}$ increases the number of systems in the synthetic sample to $\sim 425,000$. Changing the
global scale factor in this way to achieve a better match with the observed KIC does not affect the relative distribution of the stars in the
synthetic sample, but it can play a role in the target selection due to its effect on the background flux. For the following work we use the increased value of the Galaxy-wide SFR.

\subsection{Distribution in colour-colour diagram}\label{sec:colcol}

\begin{figure*}
\includegraphics[width=1.0\linewidth]{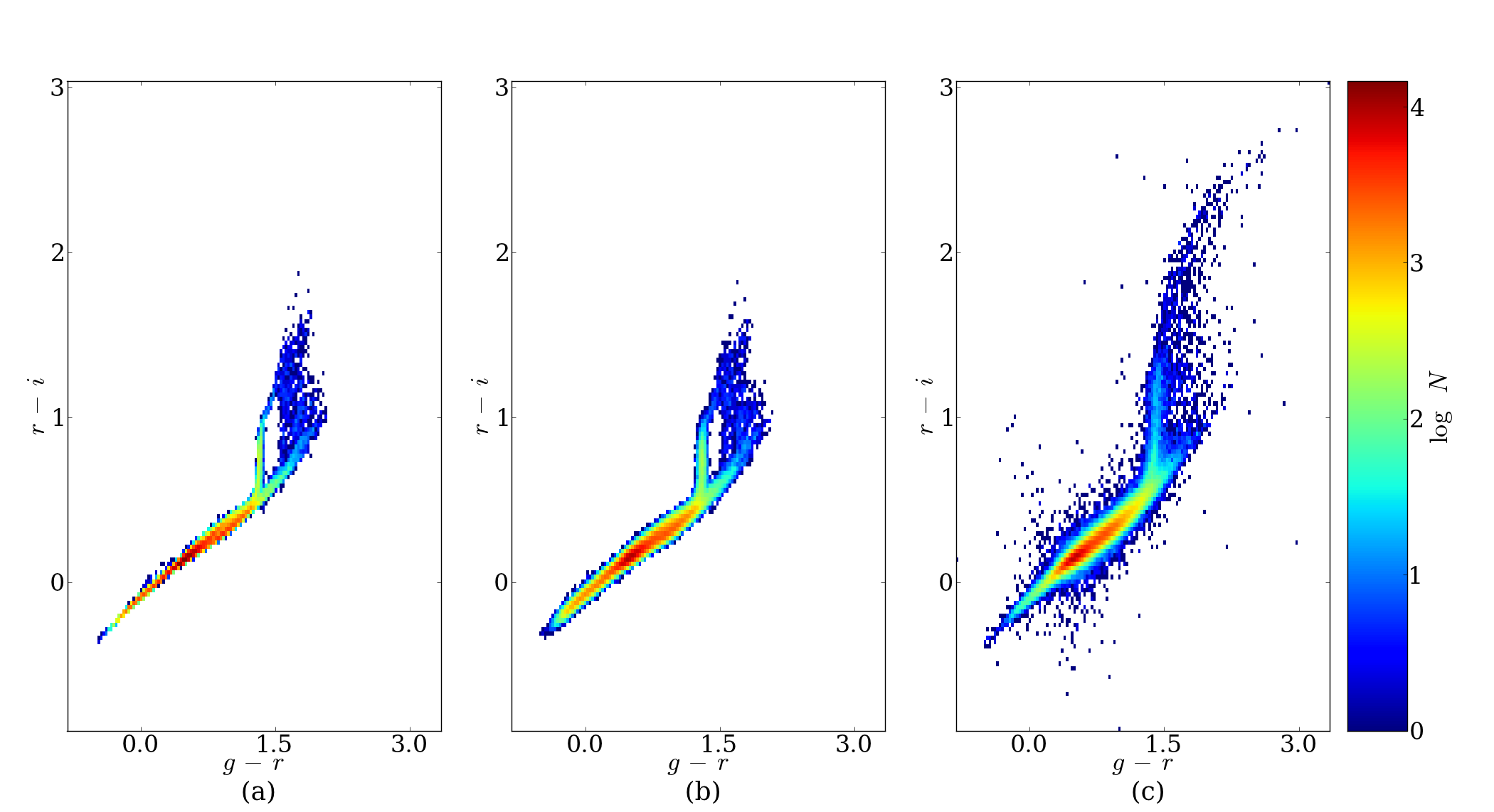}
\caption{Distribution of objects in a $r-i$ vs $g-r$ colour-colour diagram. (a) The synthetic KIC sample obtained by population models (b) the synthetic KIC sample after the application of magnitude corrections terms (c) the real KIC Q2 data set.}
\label{fig:colcol}
\end{figure*}

The distribution of KIC objects with $K_p \leq 16.0$ in the $r-i$ vs $g-r$ colour-colour diagram is shown in Fig.~\ref{fig:colcol}. The left panel shows the synthetic KIC (Fig.~\ref{fig:colcol}a) while the right panel displays the real KIC (Fig.~\ref{fig:colcol}c)
In $(g-r)$-$(r-i)$ colour space, effective temperature decreases from left to right and metallicity acts essentially perpendicular to the main band of systems, with
 higher metallicities having lower $r-i$. The fork at $g-r \sim 1.5$ is where
the dwarfs (top branch) split from the giants (lower branch), and is located at $T_{eff} \sim 3500$K. The distributions of the synthetic and real KIC display a reasonable agreement in the overall shape, however we found that when we applied
the SCP code to the synthetic sample, the resulting derived physical parameters were very sensitive to the precise location of the stars in the colour-colour diagram.

We therefore implemented a set of corrections to force a yet better agreement
between the colour-colour distributions, the result of which can be seen in the middle panel of 
Fig.~\ref{fig:colcol}. We applied a set of three
correction terms: a linear offset in each filter band, a colour-dependant
term, and a Gaussian perturbation in each filter band.
The rationale for this approach is provided
by  \citet{pinsonneault12} who found a linear offset
and a colour dependant difference term when comparing the magnitudes measured by the KIC and by the
SDSS. The Gaussian perturbation applied to all magnitudes on the other hand acts to widen the main band in the colour-colour diagram, mimicking
a more realistic, continuous metallicity distribution (rather than a bimodal one) and the effect of photometric uncertainties.

We tested the corrections
presented in \citet{pinsonneault12} to translate our SDSS based magnitudes into the KIC
based magnitudes, however these lead to unsatisfactory
fits in the resulting SCP derived parameters. This suggests that the effects of
a systematic shift between SDSS and KIC magnitudes, the simplified metallicity distribution, and superimposed photometric
uncertainties can not be separated into three independent corrections that would stand on their own.
Hence the numerical values of the corrections we derived here are particular to
our model and would not be suitable for other models, but the technique we followed may be useful to others.

To derive the corrections we applied a least squares minimisation procedure, fitting the linear offset and colour terms simultaneously, using the distributions of the synthetic and real KIC in the following colour-magnitude
diagrams: $g$ vs $(g-r)$, $r$ vs $(r-i)$, $i$ vs $(r-i)$, $z$ vs $(i-z)$, and $D51$ vs $(r-D51)$. 
For the Gaussian
terms we also used a least squares minimisation procedure to find its width for each filter band, fitting in colour-colour space.
We draw a random number from a standard normal distribution, using the same random number for
each filter, scale it by the estimated width of a Gaussian centred on the magnitude derived for the object in question,
and repeated this for each system. This was performed for the colour-colour distributions in $(g-r)$ vs $(r-i)$ and $(z-r)$ vs $(r-D51)$,
while not allowing $r$ to vary, to
derive Gaussian width coefficients for $g, i, z$ and $D51$. The procedure leads to the coefficients quoted in Table~\ref{tab:colCorr}.

Comparing the three colour-colour diagrams in Fig.~\ref{fig:colcol} we can see that the corrections
have had the desired effect. 
The agreement between the corrected synthetic sample (middle panel) and the reference sample (right panel) has improved in two important
aspects: there is a better match of the location of the peak density, and the width of the main band has also increased.
Whilst there are still some areas of improvement, for example there appear to be too many objects with $g-r \leq 0.6$ in the synthetic sample, which would translate into too
many `hot' dwarfs after target selection, and there is a lack of the reddest dwarfs, with $r-i > 1.5$, the bulk features of the synthetic sample are in satisfactory
quantitative agreement with the reference sample for the purpose of the analysis presented below.

\begin{table}
\centering
\begin{tabular}{|l|c|c|c|}
\hline
 Filter & Linear offset (l) & Colour term (c)&  $\sigma^2$ \\
 \hline
g & -0.01819 & 0.02535$(g-r)$  & 0.01921\\
r & -0.01192 & 0.05728$(r-i)$ & 0.0\\
i & -0.02209 & 0.09656$(r-i)$ & 0.00995\\
z & -0.01313 & 0.08599$(i-z)$ & 0.02611\\
D51 & -0.0222 & -0.0571$(r-D51)$ & 0.00001\\
\hline
\end{tabular}
\caption{Correction terms applied to the calculated KIC magnitudes according to
$X'_j=X_j+l+c+\sigma^2\phi$, where $X_j$ is the magnitude in filter bands $ j = g, r, i, z, D51$ and $\phi$ is a random number drawn from a standard normal distribution.}
\label{tab:colCorr}
\end{table}

\subsection{Stellar parameter distribution}\label{sec:pretarg}
Based on the corrected magnitudes we subjected all objects in the synthetic sample to analysis with the SCP code, and thus determined their `apparent' physical parameters, as obtained by the SCP. 
Thus we can compare the actual physical properties (as determined by our population model) and the SCP-derived properties of synthetic KIC stars, and check if there are significant differences between the two. By implication, we expect that any such differences would also be present in the real KIC.   
For this comparison we focus on the distribution of synthetic KIC stars in the $\log T_{eff}$ - $\log g$ diagram, as these are the most reliable parameters derived from the SCP.

\begin{figure*}
\includegraphics[width=1.0\linewidth]{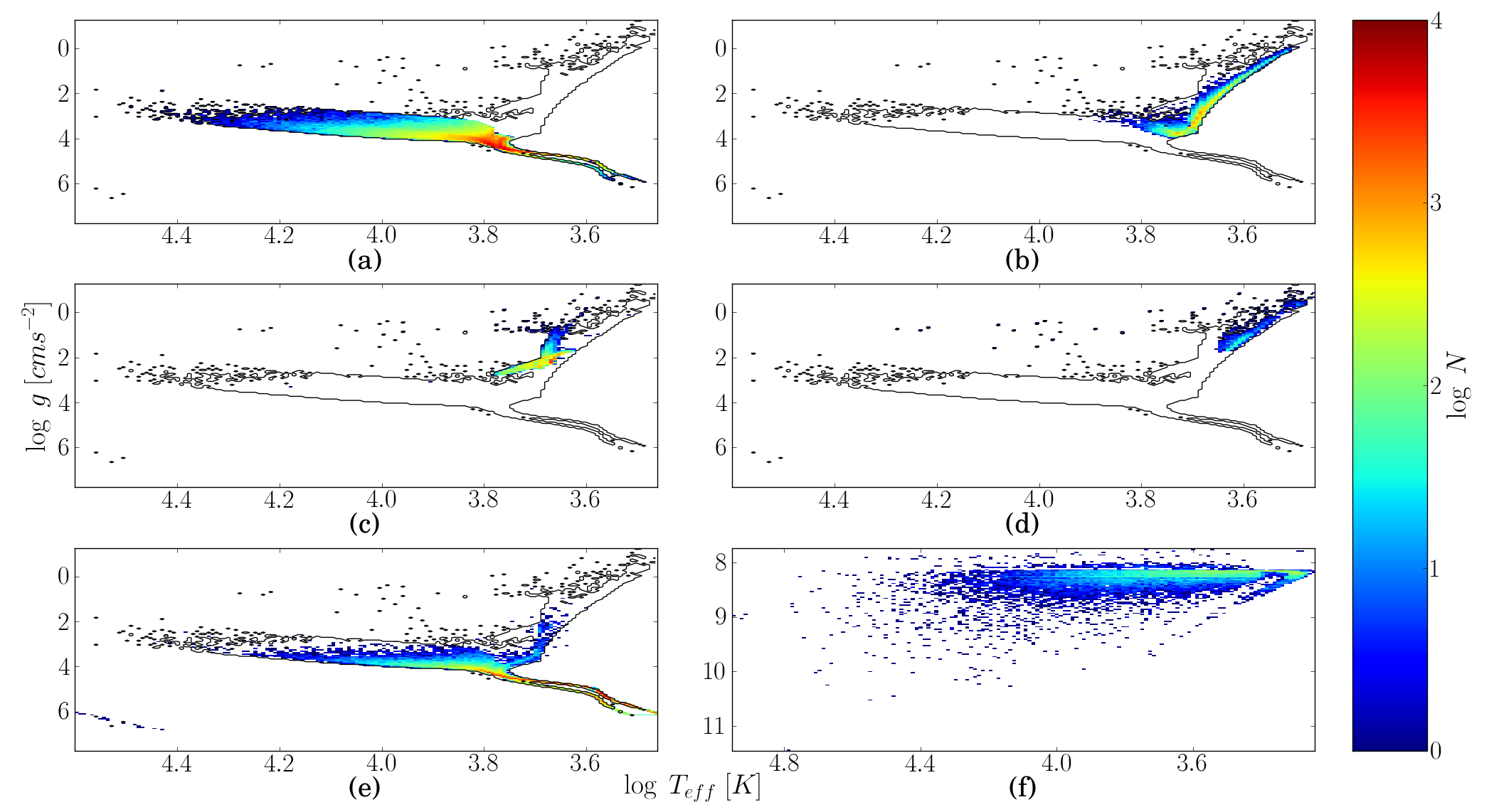}
\caption{The distribution of the synthetic KIC sample over the $\log T_{eff}$  - $\log g$ plane, for different system types. $T_{eff}$ and $g$ are the actual physical parameters of the population model stars.
In case of binaries the location of the primary is shown in panels a-d. The black contour outlines the region occupied by the combined synthetic sample.
(a) Main-sequence (MS) stars; `dwarfs'; (b) Hertzsprung gap and giant branch (hereafter GB) stars; (c) core helium burning (CHe) stars,
(d) asymptotic and thermally pulsing giant branch (AGB) stars.
(e) secondary components of a binary system, excluding systems containing a white dwarf (WD) or neutron star (NS);
(f) WDs (these are all in binaries; there are no single WDs in the synthetic sample).}
\label{fig:bisepsPre}
\end{figure*}

We first present the distribution of the actual parameters of the synthetic sample (Fig.~\ref{fig:bisepsPre}), broken up by evolutionary type. For the binaries in the sample we show the location of the primary star (except in panel e, see below).
The systems occupy a region with a bird-like shape with two prominent `wings' and  a long `neck' towards large $g$ and small $T_{eff}$. The location of this region is outlined in black in panels a-e of the figure.
The `neck' in fact consists of two narrow, essentially parallel branches which result from the bimodal metallicity distribution in our population model. The lower branch is occupied by the lower metallicity, $Z=0.0033 $, main sequence (MS) stars, while the solar metallicity MS stars are in the upper branch.
The high $T_{eff}$ `wing' is comprised of higher-mass MS stars while the other, lower $T_{eff}$ `wing' is comprised solely of evolved stars.

Panel a of Fig.~\ref{fig:bisepsPre} shows the distribution of MS stars (`dwarfs'), while panel b shows Hertzsprung gap and giant branch (GB) systems. In panel c we display core helium burning (CHe)
systems, and panel d shows asymptotic and thermally pulsing giant branch (AGB) systems.
In Fig.~\ref{fig:bisepsPre}e we show the distribution of the secondary components in
binary systems; comparing with Fig.~\ref{fig:bisepsPre}a, and in particular with the black outline, we see that in general the secondaries
are more clustered at the low $T_{eff}$, high $g$ end of the diagram. This implies that they in general have a lower mass or are less evolved than their primary companions, reflecting the fact that they were the lower mass component at birth of the binary.

Figure \ref{fig:bisepsPre}f shows the distribution of white dwarfs (WD) that are in a binary system. The synthetic sample contains no single WDs, but there are a very small number of binaries with a neutron star component (249 for the adopted input parameter set). We do not investigate the distribution of these NS systems further as our model currently treats them
in a simplistic way.

\begin{figure*}
\includegraphics[width=1.0\linewidth]{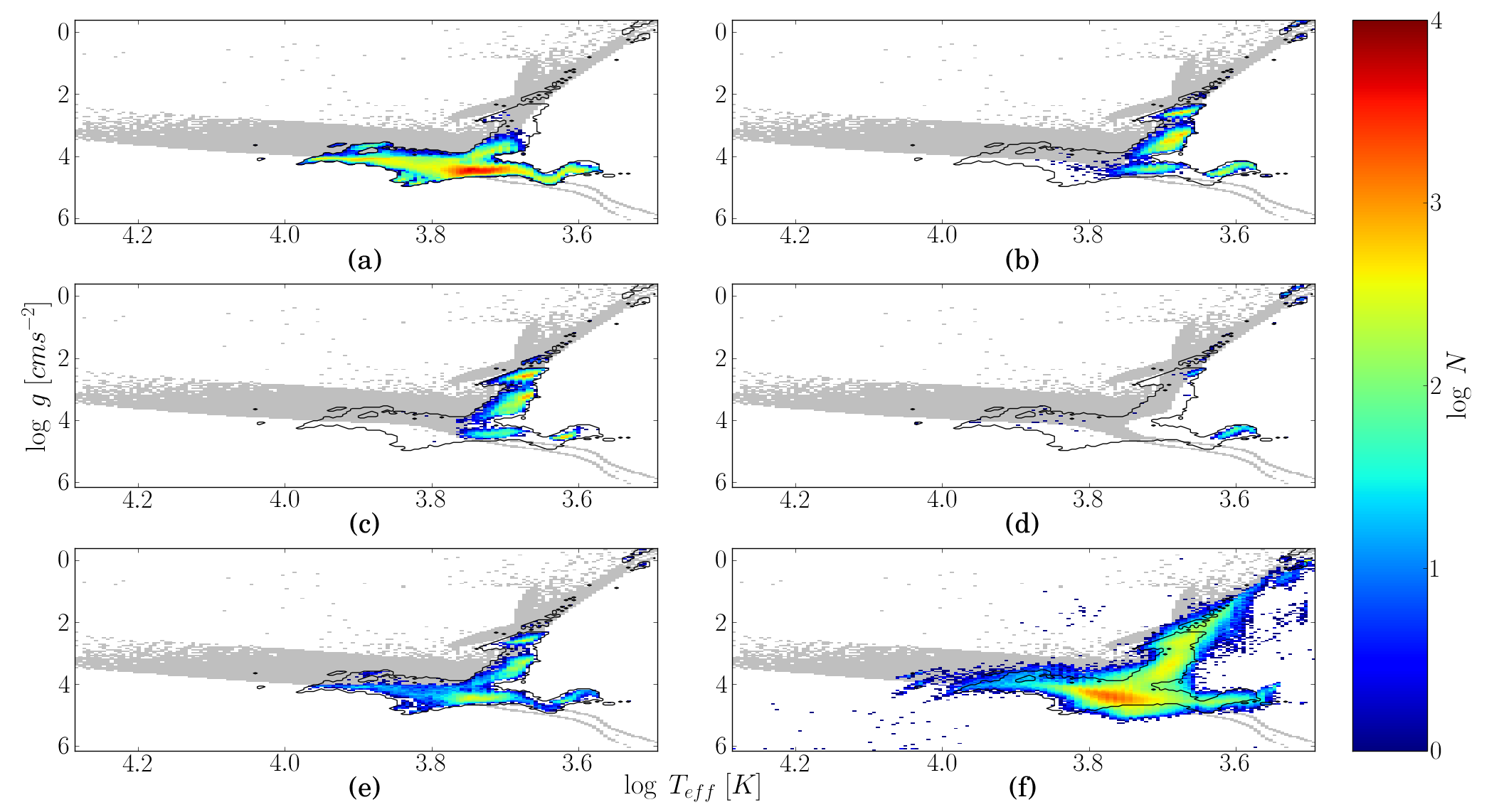}
\caption{The distribution of $\log T_{eff}$ and $\log g$ derived for the synthetic sample using
the procedure in the SCP, for different stellar types : (a) MS stars, (b) GB stars, (c) CHe stars,
(d) AGB stars and (e) systems containing a WD. 
Panel (f) shows the distribution of the Q2 data set from the SCP. 
In all panels the contour indicates the region covered by the combined synthetic distribution, while the grey-shaded area 
indicates the region covered by the distributions seen in Fig.~\ref{fig:bisepsPre}
}
\label{fig:kicPre}
\end{figure*}

We now turn to the corresponding distribution of the synthetic sample over the `apparent', SCP-determined values
for $\log T_{eff}$ and $\log g$, shown in Fig. \ref{fig:kicPre}. To aid the comparison with the previous figure a grey-shaded area indicates the region the synthetic sample occupies
in Fig.~\ref{fig:bisepsPre}.

Panels a-d in Fig.~\ref{fig:kicPre} display the same
stellar subtypes as panels a-d of Fig.~\ref{fig:bisepsPre}.
We can see that the `neck', made up of low-mass MS stars, is wider in Fig.~\ref{fig:kicPre}a than in
Fig.~\ref{fig:bisepsPre}a, and obviously is not bimodal. The `neck' is also at roughly constant $g$, while in the actual parameter space (Fig.~\ref{fig:bisepsPre}a) $g$ increases with decreasing $T_{eff}$.
The `wing' of higher-mass, more evolved MS systems (towards large $T_{eff}$) in Fig.~\ref{fig:kicPre}a is shorter than its analogue in Fig.~\ref{fig:bisepsPre}a. 
Comparing panel b in Figs.~\ref{fig:kicPre}b and Fig.~\ref{fig:bisepsPre}b reveals that giant branch stars extend over a similar range in
$T_{eff}$ and $g$, however in Fig.~\ref{fig:kicPre}b some of the giants appear at low $T_{eff}$ along the `neck', with a small gap between the bulk of the GB stars and these outliers. 
The CHe systems in Fig.~\ref{fig:kicPre}c are less constrained in $T_{eff}$-$g$ space than in
Fig.~\ref{fig:bisepsPre}c while also having a small population in the `neck'. Finally, the AGB systems
in Fig.~\ref{fig:kicPre}d appear mostly in the `neck' rather than the expected low $g$ `wing' as seen in Fig.~\ref{fig:bisepsPre}d.
The MS stars, or `dwarfs' (Fig. \ref{fig:bisepsPre}a \& \ref{fig:kicPre}a) are well constrained by the requirement
$\log g > 3.5$. However the more evolved objects (Fig. \ref{fig:bisepsPre}b-d \&
\ref{fig:kicPre}b-d) are not constrained by $\log g$ alone. So a selection based purely by $\log g$
will be able to include or exclude dwarfs, but not giants. This has ramifications for 
the bulk characteristics of the exoplanet candidate systems \citep{gaidos13}.

There is no analogous version for panel e of Fig.~\ref{fig:bisepsPre} as the SCP treats all
objects as single stars.  Instead Fig.~\ref{fig:kicPre}e shows how systems with a WD component would appear after the SCP analysis.
We find that the resulting distribution is not significantly different from systems without a WD, confirming that there is no systematic way to identify WD systems from KIC parameters alone. This lack of difference is due to the fact that the WD's luminosity is at least a factor of 100 less than its companion's luminosity, thus its flux is negligible for the colour bands that determine the solution in $T_{eff}$ and $g$.

Finally Fig.~\ref{fig:kicPre}f shows the real KIC stars (for Q2), with the black contour outlining the distribution of our SCP-processed synthetic KIC, demonstrating satisfactory agreement in terms of
overall shape and distribution. The only significant difference remaining is the lack of a continuous
giant branch track towards the lowest $g$ values.

\subsection{Post-target selection distributions}
After applying the target selection code described in Section \ref{sec:stellClass} to the synthetic population of stars
we can investigate how the target selection criteria affect the different
evolutionary types of systems compared to their intrinsic distribution.

In terms of total number of objects, 
the synthetic KIC sample was made up of 424,511 objects (208,697 single stars and 215,814 binary systems). This is reduced to 214,747 objects (104,663 single and 110,084 binaries) after target selection. The real KIC data set contains 405,789 stars while the Q2 catalogue contains 165,434 objects.
Thus the specific synthetic sample we chose to work with has 5\% more objects than the KIC to begin with, and 20\% more objects after target selection compared to the Q2 dataset. The pre-target selection number of objects could be matched perfectly by fine-tuning the underlying global Galactic SFR, but this would not affect the fraction of stars being selected as a target -  $\sim 50$\% for the synthetic vs $\sim 40$\% for the Q2 stars.

We find that the binary fraction of our sample remains largely unaltered near the 50\% level after
the application of the target selection, thus we conclude that the target selection procedure 
does not select binaries differently than it does single stars. The synthetic sample contains slightly more binaries
than single stars, due to binaries being inherently more luminous and thus a magnitude-limited sample will probe a larger volume of the Galaxy; however this difference is negligible.

\begin{table}
\centering
\begin{tabular}{|l|c|c|c|}
\hline
 Type & \multicolumn{3}{|c|}{Singles} \\
 \hline
      & Pre & Post & Relative difference\\
 \hline
MS  & 73.7\% & 79.6\% & +8.0\%\\
GB   & 15.7\% & 10.2\% & -35\%\\
CHe & 10.1\% & 9.4\% & -6.9\% \\
AGB & 0.4\%  & 0.6\% & +50\% \\
Total number & 208,697 & 104,664 & -50\% \\
\hline
\end{tabular}
\caption{ The relative distribution of stellar types among the single stars in the synthetic sample, before and after target selection.}
\label{tab:singprepost}
\end{table}

\begin{table}
\centering
\begin{tabular}{|l|c|c|c|}
\hline
 Type & \multicolumn{3}{|c|}{Binaries} \\
 \hline
      & Pre & Post & Relative difference\\
 \hline
MS  \& MS  & 68.1\% & 74.8\% & +9.8\%\\
GB  \& MS  & 11.7\% & 6.3\% & -46\%\\
WD  \& MS  & 8.0\%  & 8.5\% & +6.3\%\\
CHe \& MS  & 5.7\%  & 5.5\% & -3.5\%\\
WD  \& GB   & 3.3\%  & 1.7\% & -49\%\\
WD  \& CHe & 2.0\%  & 1.85\% & -7.5\%\\
GB  \& GB   & 0.28\% & 0.17\% & -39\%\\
Total number & 215,814 & 110,084 & -49\% \\
\hline
\end{tabular}
\caption{ The relative distribution of binary classes in the synthetic sample, before and after target selection.
Note this list has been truncated, the remaining types make up $<0.2\%$ individually and $1\%$ combined.}.
\label{tab:binprepost}
\end{table}

Tables~\ref{tab:singprepost} and \ref{tab:binprepost} show how the relative contribution from the different
stellar and binary types change after the target selection. The relative fraction of MS and MS+MS
objects increases by  $\sim 10 \%$, while the fraction of systems containing a giant decreases by $\sim 40
\%$. The original aim of the Kepler target selection was to prioritize Sun-like stars \citep{batalha10},
while also removing giant stars where Earth-sized transits are harder to detect \citep{borucki11a}.
Our analysis shows that the target selection largely succeeded in this goal, and our simulations allow one to quantify the bias this procedure introduces to the stellar sample.

The fraction of single CHe stars is almost unchanged after the target selection, most likely due
to the fact that most of them are misclassified into the dwarf region of $ \log T_{eff}$ - $\log g$ space. 
The fraction of single AGB stars increases by 50\% but is overall very small.
The CHe+MS binary systems are also unaffected by the target selection, while binaries containing an AGB star, or
CHe+GB systems, are too rare to draw conclusions from.


\subsection{Effect of target selection}

\begin{figure}
\includegraphics[width=1.0\linewidth]{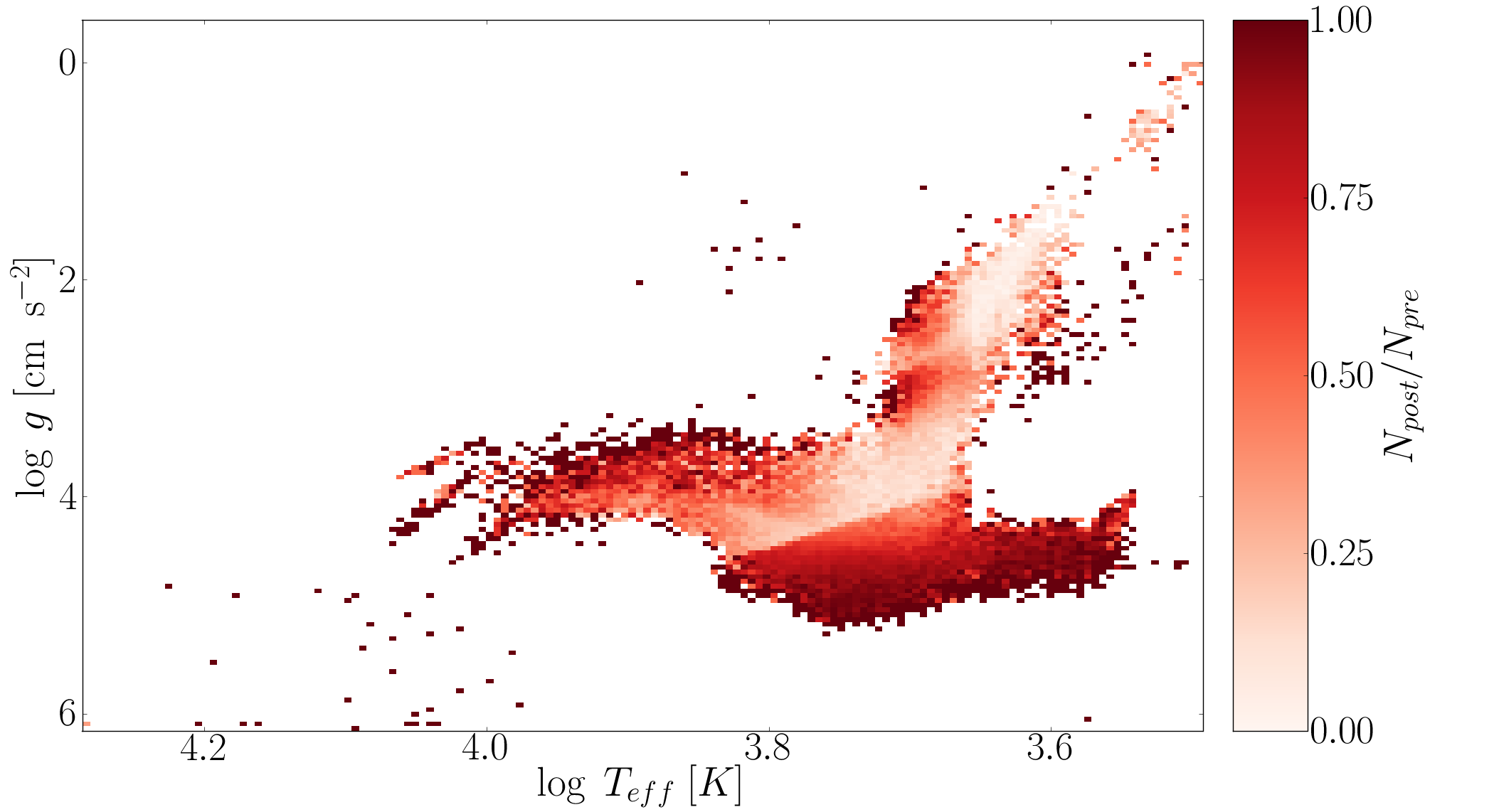}
\caption{A comparison of the distribution of systems before and after the target selection, in $\log T_{eff}$ - $\log g$ space, for the actual Q2 star sample.}
\label{fig:kickicdiff}
\end{figure}

\begin{figure}
\includegraphics[width=1.0\linewidth]{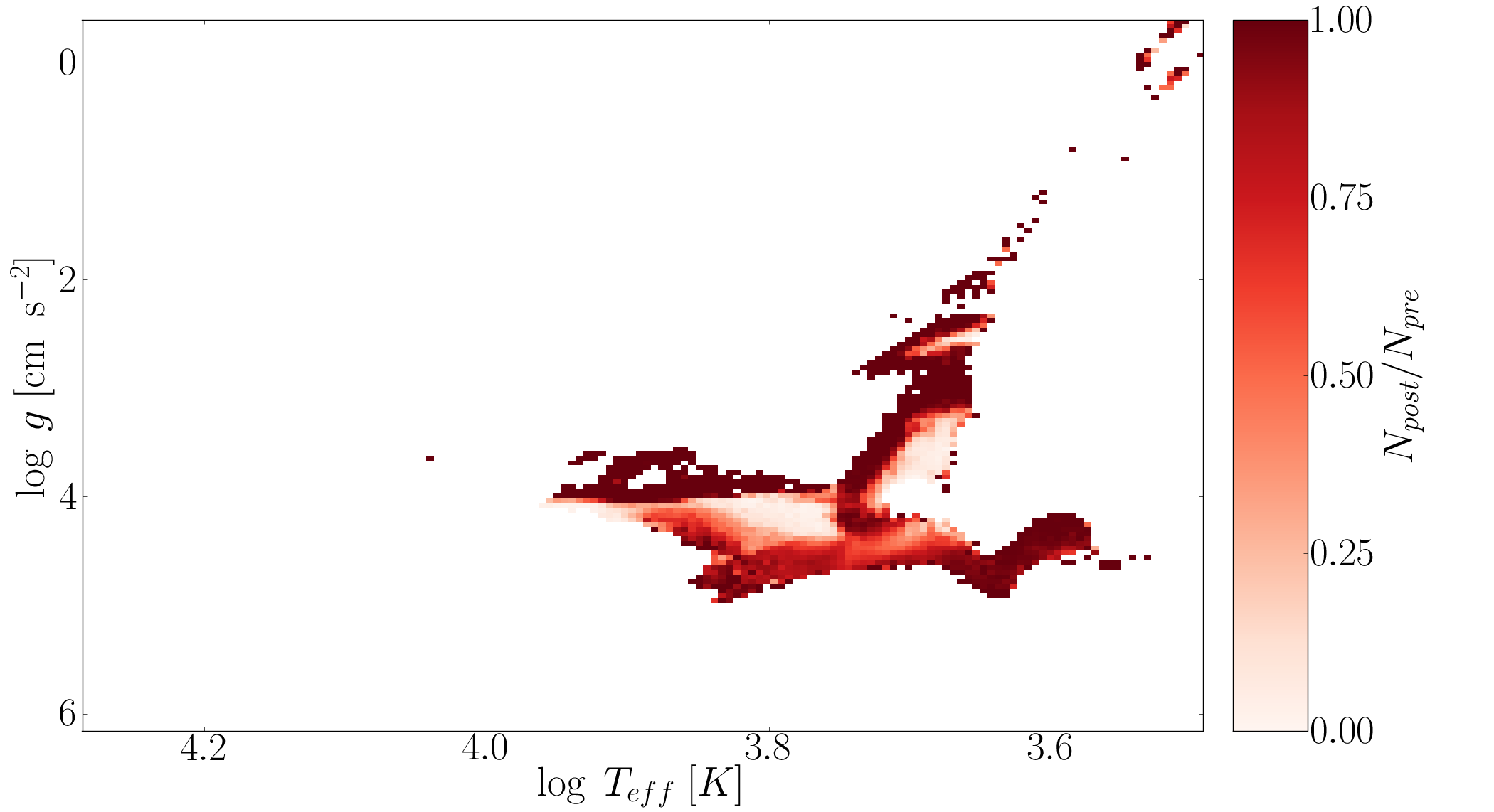}
\caption{Same as Fig.~\ref{fig:kickicdiff}, but for the synthetic sample, using SCP-derived parameters.}
\label{fig:kicdiff}
\end{figure}

\begin{figure}
\includegraphics[width=1.0\linewidth]{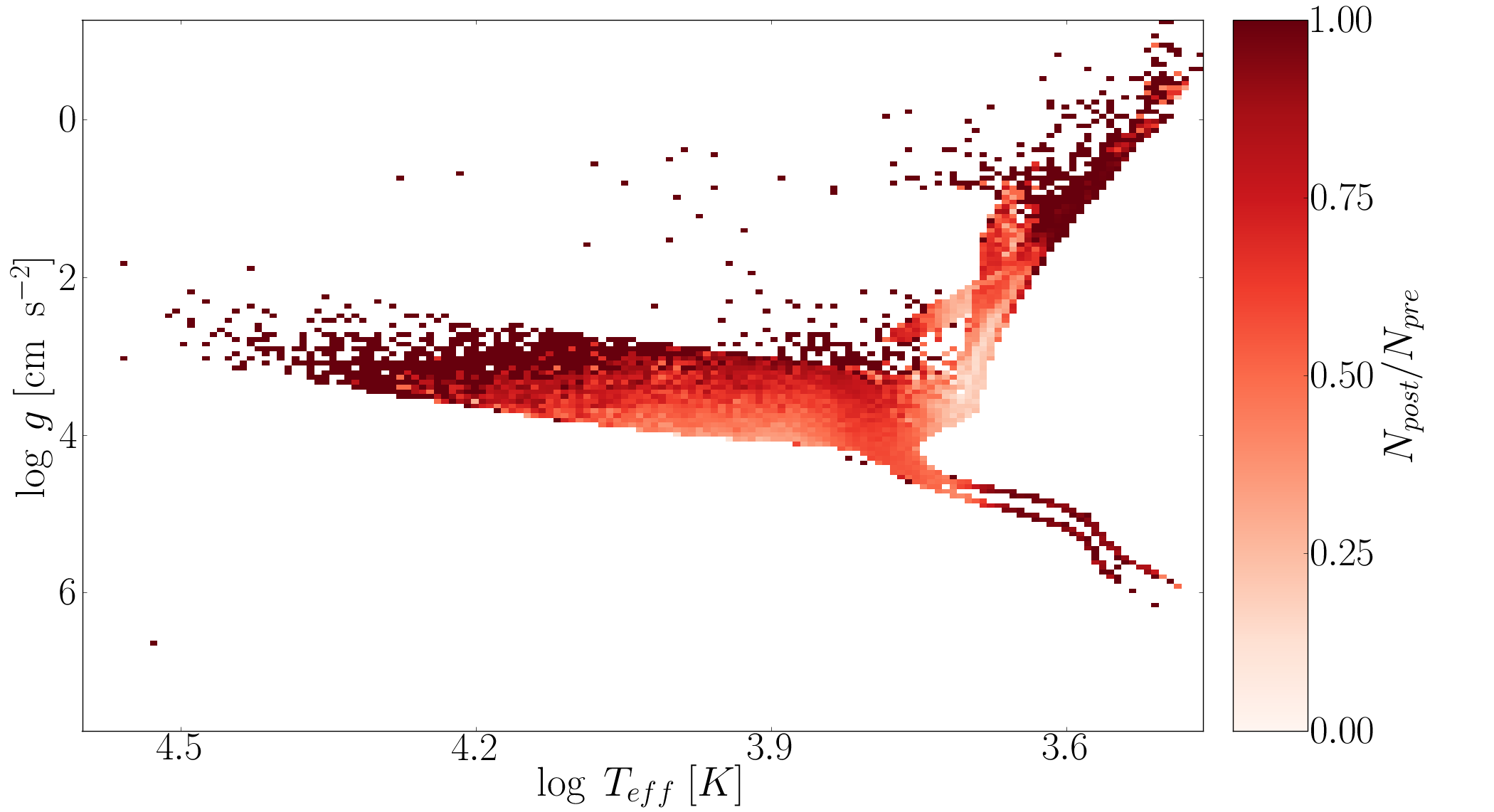}
\caption{Same as Fig.~\ref{fig:kickicdiff}, but for the synthetic sample, using their correct, physical parameters.}
\label{fig:bisepsDiff}
\end{figure}

We visualise the impact of the target selection in the $\log T_{eff}$ - $\log g$ diagram by showing the ratio of the number of systems per ($\log T_{eff}$, $\log g$) bin post- to pre-target selection, for three different samples. Figure \ref{fig:kickicdiff} compares the actual Q2 target list with the real KIC, Fig.~\ref{fig:kicdiff} considers our synthetic sample in SCP parameters, and Fig.~\ref{fig:bisepsDiff} in real parameters.

As can be seen from Fig.~\ref{fig:kickicdiff} the target selection increased the fraction of cool dwarfs and decreased the fraction of the hotter dwarfs and of giants. The change in the density between the `neck' and the `wings' is due to objects in the `neck' having $N_{tr}>3$ for objects in their HZ.

The synthetic sample in SCP-derived parameters (Fig.~\ref{fig:kicdiff})  
has a population of dwarfs in the `neck' which is comparable to those in
Fig. \ref{fig:kickicdiff}. 
The population of target-selected objects in the high $T_{eff}$ `wing' partially matches those found in Fig.~\ref{fig:kickicdiff}, though we have many more objects there. They have SCP mass $\sim1-2M_{\odot}$ and radii $\sim1.5-4R_{\odot}$, allowing the detection of a planet at $5R_{\odot}$ that would transit 3 times in 3.5 yrs. 
The giants in the low $g$ `wing' are again more marked relative to the real KIC. These are partly made up of giants that have survived the target selection criteria of \citep{batalha10} and partly due to the ad-hoc correction we applied to increase the number of objects with real radii $3<R_{\odot}<10$ (these are predominantly CHe stars).
The population of giants at the lowest $g$ values is due to the ad-hoc correction that adds objects that saturate at least one pixel.

Figure \ref{fig:bisepsDiff} finally reveals how the synthetic sample is target-selected as a function of actual, physical parameters. The population of dwarfs in the `neck' of Fig.~\ref{fig:bisepsDiff} matches well with the population in the `neck' of Fig.~\ref{fig:kicdiff}. The `hot' dwarfs are still present in Fig. \ref{fig:bisepsDiff}. 
Note that the large number of target-selected objects in the GB and AGB `wing' are due to their misclassification by the SCP (objects seen in Fig.~\ref{fig:kicPre}b-d in the `neck'). They have SCP-derived $\log g$ values of 4.2-4.6 which implies an SCP-derived mass $M=0.5-0.8M_{\odot}$; hence these objects were in fact classified into our highest priority target group. 
The overpopulation of giants noted in Fig.~\ref{fig:kicdiff} is less pronounced in Fig.~\ref{fig:bisepsDiff}, but here they reside in the CHe region (see Fig.~\ref{fig:bisepsPre}c) and the extreme end of the AGB region (see Fig.~\ref{fig:bisepsPre}d).

\subsection{Comparison of SCP and physical parameters}
For a closer inspection of the differences between the real, physical parameters and the SCP-derived parameters we compare the synthetic sample and the KIC before target selection, as this maximises the number of objects to derive results from.  For each object in the sample we determine the difference between the real and SCP-derived effective temperature,  $\Delta T_{eff}= T_{eff, real}-T_{eff, SCP}$, and surface gravity, $\Delta \log g= \log g_{real}-\log g_{SCP}$. In case of a binary system only the primary star is considered.
We then adopt a suitable binning of the $\log T_{eff}$ -$ \log g$ plane and determine, for each bin, the median values of $\Delta T_{eff}$ and $\Delta \log g$ for all objects that fall into a given bin.

\begin{figure}
\includegraphics[width=1.0\linewidth]{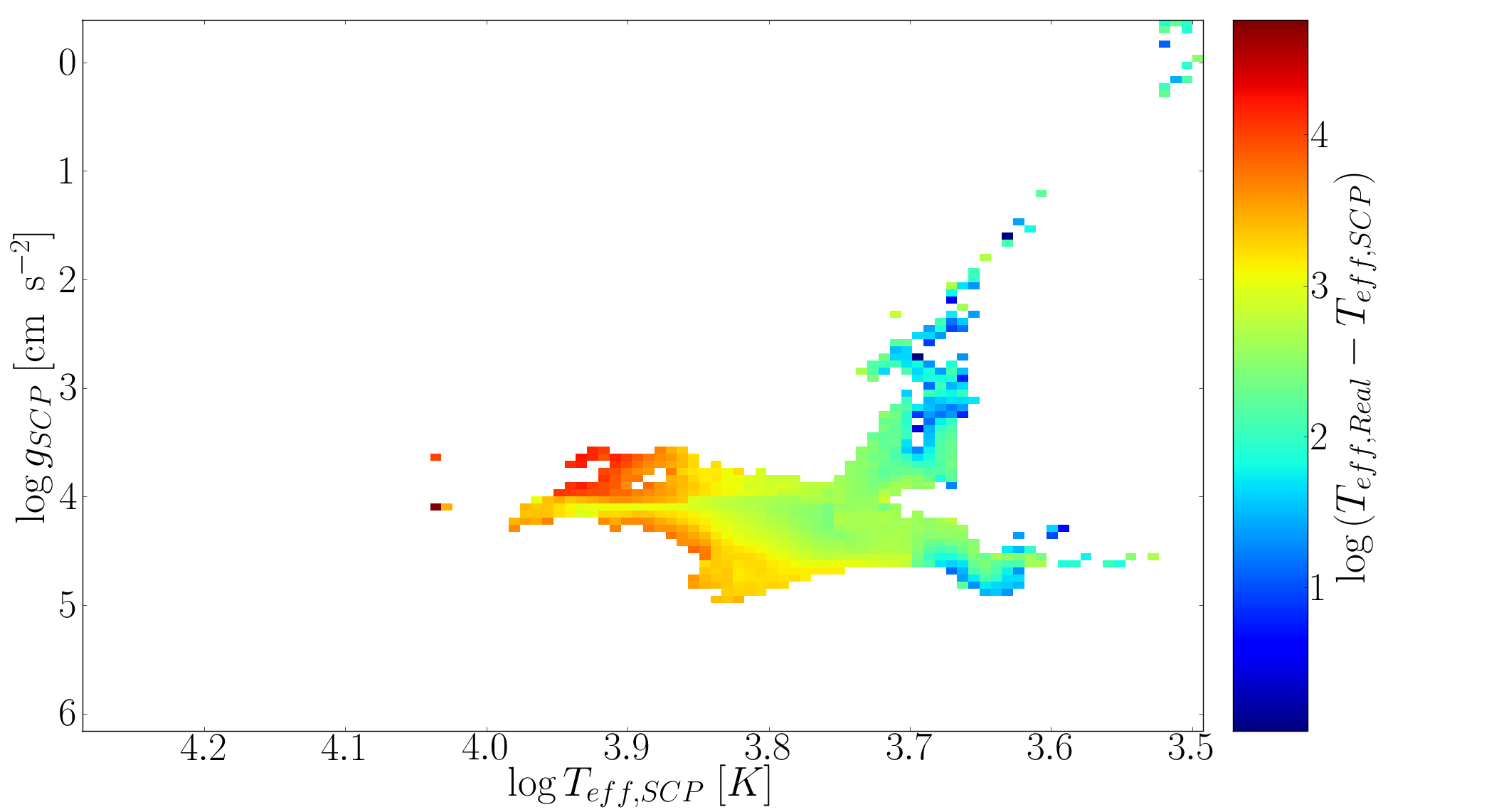}
\caption{ Distribution of the median of the difference between real and SCP-derived effective temperature, $\Delta T_{eff}= T_{eff,real}-T_{eff,SCP}$, per bin of $\log T_{eff}$ - $\log g$.}
\label{fig:teffdiff}
\end{figure}

\begin{figure}
\includegraphics[width=1.0\linewidth]{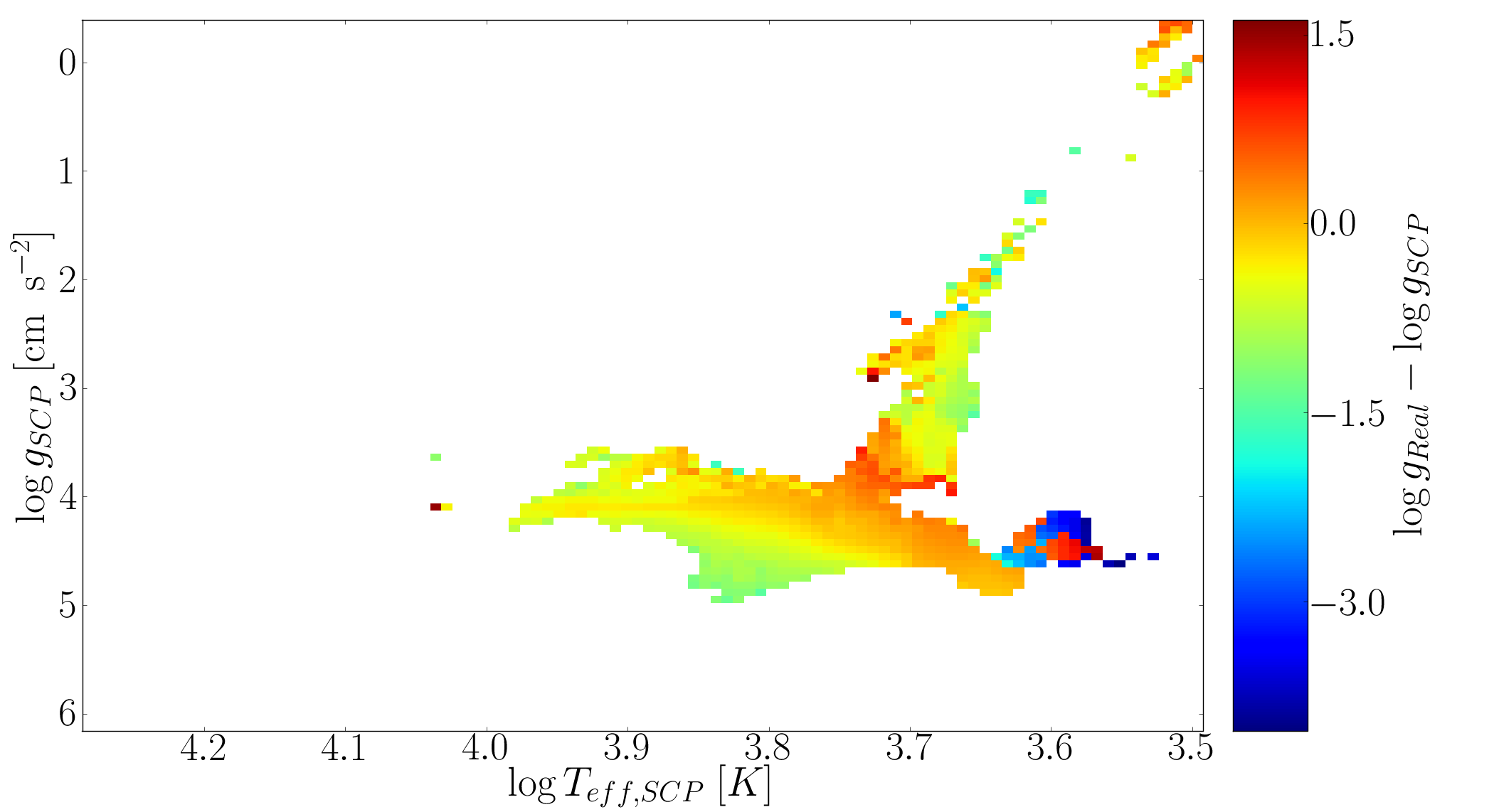}
\caption{Distribution of the median of the difference between real and SCP-derived surface gravity, $\Delta \log g= \log g_{real}-\log g_{SCP}$, per bin of $\log T_{eff}$-$\log g$.}
\label{fig:loggdiff}
\end{figure}

Figure \ref{fig:teffdiff} shows $\Delta T_{eff}$ as a function of $\log T_{eff}$ and $ \log g$. The largest differences are seen in the hottest dwarfs. This is not unexpected as the SCP had a $T_{eff}$ limit of $50,000$~K \citep{brown11}. 
Figure \ref{fig:loggdiff} displays the
distribution of $\Delta \log g$ over the $\log T_{eff}$ - $ \log g$ plane. 
The population of giants that are in the `neck' and misclassified as dwarfs are clearly visible, having the largest $\Delta \log g$. 

Tables \ref{tab:deltaSing} and \ref{tab:deltaBin} show the median values of $\Delta T_{eff}$ and $\Delta \log g$ across the whole parameter space, and the corresponding standard deviation, $\sigma$, binned on evolutionary type.  MS systems (MS, MS+MS \& WD+MS) have the largest values of $\Delta T_{eff}\sim 500$~K as well as the largest standard deviations, which is caused by the hot dwarfs. The evolved systems (GB, CHe \& AGB-containing systems) all have relatively small values, $\Delta T_{eff} <100$~K. 


Comparing Tables~\ref{tab:deltaSing} and \ref{tab:deltaBin} we conclude that the SCP method performs in a similar way on binaries as it does for single stars.

On average, the SCP-derived value for $g$ is smaller than the real value. The SCP will therefore return a larger radius than the real radius, and consequently any derived planet radius will be larger as well. 
If $\log g$ for an MS star is underestimated by the average value of 0.23 dex the implied stellar radius is too large by $\sim3\%$. For a measured transit depth $\Delta F/F= (R_p/R_*)^2$ the planet radius$R_p$ will also be overestimated by $\sim3\%$, and its bulk density underestimated by nearly $10\%$ if the stellar mass is assumed known. While confirmed Kepler planets will have stellar radii determined by other means, usually by spectroscopy \citep{batalha11}, most systems are too faint, and they are too numerous, for affordable, individual follow up \citep{batalha10}, thus their radii will be uncorrected in the first instance and any derived planetary distributions skewed.

\begin{table}
\centering
\begin{tabular}{|l|c|c|c||c|c|}
\hline
 Type & \multicolumn{5}{|c|}{Singles} \\
 \hline
 & \multicolumn{2}{|c|}{$\Delta T_{eff}\; [\rmn{K}]$}& & \multicolumn{2}{|c|}{$\Delta \log g\; [\rmn{dex}]$} \\
\cline{2-3} \cline{5-6}
   & Median & $\sigma$ & & Median & $\sigma$ \\
 \hline
MS    & 492 & 918 & & -0.23 & 0.39 \\
GB    & 61 & 197 &  &-0.42 & 0.96 \\   
CHe  & 74 & 214 &  &-1.01 & 0.67 \\
AGB & -23 & 3758 & & -3.01 & 1.14 \\
\hline
\end{tabular}
\caption{The median values of the differences $\Delta T_{eff}$ and $\Delta \log g$ (with the corresponding standard deviation $\sigma$), between the real, physical parameters and the SCP-derived parameters for our synthetic single stars.}
\label{tab:deltaSing}
\end{table}

\begin{table}
\centering
\begin{tabular}{|l|c|c|c|c|c|c|}
\hline
 Type & \multicolumn{5}{|c|}{Binaries} \\
\hline
 & \multicolumn{2}{|c|}{$\Delta T_{eff}\; [\rmn{K}]$}& & \multicolumn{2}{|c|}{$\Delta \log g\; [\rmn{dex}]$} \\
\cline{2-3} \cline{5-6}
   & Median & $\sigma$ & & Median & $\sigma$ \\
 \hline
MS  \& MS  & 558 & 931 & & -0.24 & 0.41\\
GB  \& MS  & 56 & 268 & & -0.48 & 0.72\\
WD  \& MS   & 471 & 615 & & -0.29 & 0.38\\
CHe \& MS  & 58 & 229 & & -1.06 & 0.65 \\
WD  \& GB  & 53 & 175 & & -0.47 & 0.78 \\
WD  \& CHe  & -6 & 708 & & -3.10 & 0.98 \\
GB  \& GB   & -8 & 258 & & -0.5 & 0.82 \\
\hline
\end{tabular}
\caption{Same as Table~\ref{tab:deltaSing}, but for binaries and only considering the primary star.}
\label{tab:deltaBin}
\end{table}

Other authors find similar results. In the SCP paper, \citet{brown11} compared the KIC estimates for some 35 stars with spectroscopic measurements, and noted for dwarfs with $T_{eff}=4500-6500$K a temperature difference $\Delta T_{eff}=\pm200$K and surface gravity difference $\Delta \log g=-0.4$~dex. 
Sampling our synthetic dwarfs over this $T_{eff}$ range we find a median value of
$\Delta T_{eff}=+423$K with $\sigma=231$K and $\Delta \log g=-0.14$~dex, $\sigma=0.33$~dex. 

\citet{pinsonneault12} modelled SDSS stars in the Kepler field with the infra red flux method (IRFM) and derived an average deviation of $\Delta T_{eff}=+215\pm 100$K for dwarf stars between 4000\ K and 6500\ K. Taking our
systems over a similar temperature range and only considering dwarfs we find a median $\Delta
 T_{eff}=+413$K, $\sigma=240$K, consistent with \citet{pinsonneault12}. 

\citet{mann12} found from medium-resolution spectra of 382 stars, $\Delta T_{eff}=-110^{+15}_{-35}$K for dwarfs and $\Delta T_{eff}=-150^{+10}_{-35}$K for giants.
Following their selection of objects with $K_p - J > 2.0$ our synthetic sample gives $\Delta T_{eff}=-140$K, $\sigma=116$K for dwarfs and $\Delta T_{eff}=+44$K, $\sigma=213$K for giants, consistent with \citet{mann12}.

\citet{dressing13} used a set of Dartmouth stellar evolution and atmosphere models, with $M<1M_{\odot}$ and $T_{eff}<7000$K, to model 3897 KIC objects with $T_{eff,KIC}<4000$K and thus derive from the KIC photometry improved stellar parameters. They found for a typical cool star in their sample that the temperature is cooler than the one quoted in the KIC, $\Delta T_{eff}=-130$K,  and that the radius is 69\% of the KIC radius. Applying their selection criteria to our population model we reproduce this result with the synthetic sample, obtaining $\Delta T_{eff}=-180$K, $\sigma=200$K, and on average a radius that is 62\% (with $\sigma=27$\%) of the KIC radius.

The apparent differences between the various authors can be naturally explained as each author focuses on systems with different temperature and different selection criteria, thus sampling different regions of the parameter space.

\section{Discussion}\label{sec:discuss}

We arrived at a synthetic model of the KIC and of the corresponding target-selected subsample by adapting a full stellar and binary star population synthesis model to the specific circumstances of the Kepler field and the Kepler detector. In order to do so we necessarily had to adopt a number of simplifications and ad-hoc assumptions. Here we discuss the potential impact these may have on our results, and what improvements further work should consider. 

The lack of a realistic metallicity distribution, we believe, is the most limiting simplification of our model. The current design of our population synthesis procedure makes the inclusion of an initial metallicity that continuously varies with Galactic epoch computationally too expensive. The adopted bimodal model highlights the variation with $Z$, but does neither span the full range of metallicities implied by SCP fits, nor cover the bracketed range in a continuous fashion. To mimic the effect of a continuous metallicity distribution we had to introduce small, random perturbations of the calculated stellar magnitudes. This approach however cannot fully capture the effect of metallicity on evolutionary timescales and system appearance --- metal-poor stars have a shorter MS life and are less luminous than stars with solar metallicity, ultimately resulting in differences in their respective distributions in the colour-colour diagrams we set out to match. To test the significance of the single metallicity value we used for the thin disc, $Z=0.02$, we modelled a small FOV corresponding to one Kepler CCD channel (roughly 6500 systems)
with a thin disc metallicity of $Z=0.014$, \citep{asplund09,nordstrom04}. We found no statistically significant differences in the stellar parameter distribution, i.e.\ the differences were smaller than those seen by random sampling with 6500 systems of the underlying probability distribution function $\Gamma$. 

Changing the metallicity alters the colour distribution in Fig.\ref{fig:colcol}. Lowering the metallicity leads in general to lower effective temperatures, thus shifts the stellar flux to longer wavelengths. In Fig.\ref{fig:colcol}a, objects with $g-r < 1.5$ are shifted upwards by $~0.02$, towards larger $r-i$, and systems with $g-r > 1.5$ are shifted rightwards, to larger $g-r$ by $~0.2$.

However, as discussed above, to force a better agreement between the synthetic sample and the real KIC we apply a series of colour-correction terms to the synthetic stars. The colour distribution for the lower thin disc metallicity will therefore
simply result in a slightly different set of corrections to achieve the best fit to the actual KIC distribution seen in Fig.\ref{fig:colcol}c, thus effectively eliminating the underlying differences. We conclude that our results are not sensitive to the choice of single-value metallicity in the thin disc.

In this context we note that the SCP itself is inconsistent in its use of metallicity. In assigning a metallicity to a given object the SCP disregards the metallicity from the stellar input models (\citet{castelli04} and \citet{girardi00}) and exclusively relies on solar metallicity ($Z=0.02$) models \citep{brown11}.

We note that the resulting SCP-derived parameters of the synthetic stars are sensitive to the colour corrections, so great care has to be taken not to introduce spurious features into the synthetic distributions. We expect that the introduction of a realistic metallicity distribution, whilst keeping a Gaussian perturbation approach to model photometric uncertainties, would reduce these corrections to a term 
dependant on the difference between the SDSS magnitudes and the KIC filter system. Such a term can then be independently constrained by e.g. \citet{pinsonneault12}. 



For the purpose of the current study we chose to keep commonly used population parameters fixed. There is considerable uncertainty in some of them, and we will present  a systematic study of their impact on the general properties of the synthetic sample in a separate paper. 
We will however briefly discuss our choice and the impact of some of these parameters.

We tested the sensitivity of our model to the assumed Galactic distribution of stars by decreasing all scale heights and radial scale lengths by 25\%, and recalculating the synthetic sample for one CCD channel. This has the effect that for a given line of sight we now sample stars that are `further' away (in units of scale lengths) from the Galactic centre, so the proportion of old disc stars, which are predominately low mass MS and WD systems, increases from 14\% to 20\%. After re-normalising the stellar density, to remain consistent with the KIC number count, we however find this effect to be negligible for the Kepler field, with only small differences, of order the random sampling noise,from the results for the original scales lengths.
The length scale reduction considered in this test is large compared to the range proposed in the recent literature. In particular, 
\citet{juric08} found scale heights that are consistent with our adopted values, except their thick disc scale height is $10\%$ smaller. In fact, using their values, we also find no significant differences to the results presented here.

The total binary fraction is, somewhat arbitrarily, set at 50\%. This allows us to study the differential effect of the SCP and the target selection on the binary content in general. In reality the binary fraction is likely to be a function of stellar mass, reaching values near 100\% for high-mass stars. The choice of binary fraction becomes a more important concern when considering Kepler's sample of eclipsing binaries or the false positive transit signal. 

The high-mass end of the IMF is well constrained, and indeed the overall normalisation of the SFR is based on this. However, the shape of the IMF below $\sim 0.5$~M$_\odot$ is more uncertain, and we have indeed tested if this offered a way to boost the number of faint objects in the target-selected synthetic sample. We found that varying the low-mass IMF within physically reasonable limits does not significantly change the magnitude distribution of the synthetic sample.

The adopted flat initial mass ratio distribution (IMRD) is preferred by many population synthesis studies, including those by \citet{girardi05} (but note that these authors add binaries to the population of single stars in an ad-hoc way, while our model treats evolving binaries self-consistently). We assessed the effect of the IMRD by re-calculating a Kepler subfield population with an IMRD that favours equal mass companions ($n(q) \propto q$) and one that favours unequal mass ratios ($n(q) \propto q^{-1}$); $q$ is the mass ratio.
 
In the former case the population of systems with near-identical components increases. The SCP will assign the correct stellar parameters of one component to the combined object, so the net effect is that a magnitude-limited sample such as the KIC will include relatively more of these objects as they are intrinsically brighter and hence can be seen out to larger distances. The apparent binary fraction does indeed increase to $\sim 55\%$ (it was $\sim 50\%$ in our standard model with a flat IMRD) in the KIC sample, and remains unchanged after target selection.

In the case of favouring unequal mass ratios on the other hand the SCP will pick out the correct parameters for the primary, but the apparent binary fraction is only $\sim3\%$. This is because favouring unequal component masses requires high-mass stars which are rare overall.

In both cases the apparent binary fraction remains unchanged after target selection.
After correcting for the change in the total number of systems, to obtain the appropriate background flux,  we find that after target selection the two different IMRDs increase (decrease) the fraction of dwarfs (giants), by $\sim5$pp, in the same way as for the flat IMRD
(see Tables~\ref{tab:singprepost} and \ref{tab:binprepost}). We also found that the median $\Delta T_{eff}$ and $\Delta \log g$ are the same as for the flat IMRD, within the quoted uncertainties.

In our population model we have ignored the fact that binaries form with eccentric orbits and circularise on a finite, system-dependent timescale. Instead we kept binary orbits circular at all times. This seems justified as \citet{hurley02} showed that the circularisation
time-scale for interacting binaries is short enough to not alter bulk properties of the binary population from one where the eccentricity is kept at 0. There is also no suggestion that the eccentricity of a binary orbit would have any effect on the system's detectability in the Kepler field.  

The adopted Galactic absorption model obviously affects the make-up of the synthetic KIC, but with the smallest scale modelled by \citet{drimmel03} being $0.35^o \times 0.35^o$ we deem this well suited to resolve the statistics of the larger Kepler field. We note that the 
Kepler team assumes a smooth, exponentially decaying absorbing disk \citep{brown11} which on average returns a larger extinction for a given distance than \citet{drimmel03}. The Kepler team quote that most of the target-selected stars are within 1~kpc from the Sun, with 
$\sim 50$\% of objects suffering a V band extinction $A_V<0.4$.  In contrast, in our model only 30\% of the target-selected stars are at $<1$~kpc, while 70\% are at $<2$~kpc, which also corresponds to $A_V<0.4$. Using the Kepler extinction model reduces the number of objects seen with $Kp<16$ by $\sim5\%$, while leaving the underlying distributions the same, within the limits of random sampling.

We will address the impact of these population parameters on the synthetic Kepler field in a separate study, where we will attempt to extract constraints on the binary fraction and initial distribution functions from the observed eclipsing binaries in the Kepler field, and from the statistics of the rare cases of binaries that show an asteroseismological signal from both components.

In order to accomplish a proper treatment of a variable initial metallicity, or to explicitly take into account initial eccentricity distributions, or additional distributions of physical parameters characterising the stellar population such as stellar rotation rates, we have to alter 
the central concept of our population code. For this task we need to switch from 
sampling of the Galactic distribution function $\Gamma$, introduced in Sec.~\ref{sec:subsample}, 
to a Monte-Carlo sampling of the initial distribution functions, then only evolving those
systems we have sampled. Depending on the desired application this may reduce the number of systems to be evolved, and we can replace the analytic fits currently used by BiSEPS to describe stellar evolution with a numerical, state-of-the-art 1D stellar evolution model, like MESA \citep{paxton11}.

\section{Conclusions}\label{sec:con}

In this work we presented a comprehensive population synthesis model of the Kepler field, taking into account single and binary star evolution. We have also modelled the selection effects inherent in the Kepler objects of interest, the SCP parameter estimation, Kepler's instrumental noise and the targeted selection of systems with the highest chance of detecting an Earth-like planet round a Sun-like star in the HZ. 
The main output of this procedure is a synthetic catalogue of systems in the Kepler field. This catalogue was the basis for a comparison between the real physical parameters of the catalogue stars, as indicated by the population model, and the corresponding SCP-derived parameters. Such a comparison over the bulk of the Kepler field is only possible with a full theoretical population model; purely observational tests of the SCP performance will always be limited to a small sample of stars on the basis of bespoke spectral fitting.
Using the synthetic sample we also investigated the effect of the target selection method on the underlying distributions in both SCP and real parameter space.

We found satisfactory agreement between the synthetic KIC and the real KIC in colour-colour space, and between our target selection method and the Q2 target selection. Our simulations highlight a difference between the physical parameters of the stars in the synthetic sample and those derived by the SCP for the synthetic sample. We conclude that this systematic difference does also exist for the SCP-derived parameters of the objects in the real KIC. Specifically, for systems containing a MS star, the SCP-derived parameters deviate on average by $\sim\Delta T_{eff}=500$~K and $\sim \Delta \log g=-0.2$~dex from the real physical parameters. In case of GB stars the deviation is $\sim \Delta T_{eff}=50$~K and $\sim\Delta \log g=-0.5$~dex. This has the remarkable consequence that the SCP-derived stellar radii of MS stars are on average too large by $\sim3\%$. If these radii are used to estimate the radius of any planet observed to be transiting then the planet radius will be $\sim3\%$ too large.

After correcting for selection effects we find that these results are consistent with differences highlighted by other authors, on the basis of observational consideration of subsamples. The average deviation for a given stellar type is observed regardless of if the star is single or in a binary. 

Our models confirm that the Kepler target selection procedure increases the fraction of main-sequence stars, from about 75\% to 80\%,  and decreases the fraction of giants, from 25\% to 20\%, relative to the KIC. In fact, our population synthesis approach is the only way to  quantify this bias; the figures demonstrate that the change is only moderate.

The bias introduced into the target-selected sample is roughly the same for single stars and binary systems. 
We also found that the target selection has a negligible effect on the binary fraction, and that it does not alter the relative fractions of systems with different stellar evolution types, when compared to the single star population.

The techniques presented here will be used in a future study to interpret the binary sample observed by Kepler, and to re-assess the Kepler false positive rate. 

\bibliographystyle{mn2e_2}
\bibliography{kepler_v15}
\section*{Acknowledgements}
R.F acknowledges support from a STFC studentship. This paper includes data collected by the Kepler mission. Funding for
the Kepler mission is provided by the NASA Science Mission directorate.
Some/all of the data presented in this paper were obtained from the Mikulski Archive for Space Telescopes
(MAST). STScI is operated by the Association of Universities for Research in Astronomy, Inc., under NASA

\end{document}